\documentclass[twocolumn,superscriptaddress,aps,preprintnumbers,amsmath,amssymb,prl,nofootinbib]{revtex4}

\usepackage{graphicx}
\usepackage{epstopdf}
\usepackage{dcolumn}
\usepackage{bm}
\usepackage{hyperref}
\usepackage{color}

\usepackage{amsmath}

\def\dd{\mathrm{d}}

\def \F{\mathcal{F}}

\begin{document}

\title{
Optical Ring Cavity Search for Axion Dark Matter
}

\author{Ippei Obata}
\affiliation{Institute for Cosmic Ray Research, University of Tokyo, Kashiwa 277-8582, Japan}
\author{Tomohiro Fujita}
\affiliation{Department of Physics, Kyoto University, Kyoto, 606-8502, Japan}
\author{Yuta Michimura}
\affiliation{Department of Physics, University of Tokyo, Bunkyo, Tokyo 113-0033, Japan}

\begin{abstract}
We propose a novel experiment to search for axion dark matter which differentiates the phase velocities of the left and right-handed polarized photons.
Our optical cavity measures the difference of the resonant frequencies between two circular-polarizations of the laser beam.
The design of our cavity adopts double-pass configuration to realize a null experiment and give a high common mode rejection of environmental disturbances.
 We estimate the potential sensitivity to the axion-photon coupling constant $g_{a\gamma}$ for the axion mass $m \lesssim 10^{-10}$ eV.
 In a low mass range $m \lesssim 10^{-15}$ eV, we can achieve  $g_{a\gamma} \lesssim 3\times 10^{-16} ~\text{GeV}^{-1}$ which is beyond the current bound by several orders of magnitude.
\end{abstract}
\maketitle

%
%
%
\section{I. Introduction}
Axion is one of the most known particles in the extended theories beyond the standard model of particle physics. It was originally introduced by Peccei and Quinn to solve the strong CP problem in QCD~\cite{Peccei:1977hh}. 
Moreover, string theory and supergravity generically predict a plenitude of axion-like particles (ALPs) which can have a broad range of the mass~\cite{Svrcek:2006yi}.
Especially, the axion with a small mass $m \ll 1\text{eV}$ is well motivated by cosmology, since 
it behaves as a non-relativistic matter fluid in 
our universe and is a good candidate for dark matter.
Axion dark matter may weakly interact with known standard model particles, so that we can explore axion dark matter through the direct search experiments.

It has been known that if axion is coupled to photon, the axion-photon
conversion under static magnetic fields takes place~\cite{Sikivie:1983ip}.
Making use of this conversion process, many different types of experiments have been considered such as axion haloscopes  \cite{Hagmann:1998cb}, axion helioscopes \cite{Zioutas:2004hi,Vogel:2013bta}, ``light shining through a wall" experiments \cite{Ehret:2009sq, Betz:2013dza}, laser interferometry \cite{Tam:2011kw, DeRocco:2018jwe} and magnetometers \cite{Sikivie:2013laa, Kahn:2016aff} (for more details, see recent reviews \cite{Graham:2015ouw} and references therein). They put the constraints on the photon-axion coupling constant $g_{a\gamma}$ for a vast range of the ALPs mass.
Astronomical observations can be also used to probe the axion-photon conversion.
For the low mass range, the absence of gamma-ray emission from SN1987A \cite{Brockway:1996yr, Payez:2014xsa}
and the spectral of cosmic rays from galaxy clusters~\cite{Wouters:2012qd, Marsh:2017yvc, Conlon:2017ofb} are used to put significant bounds on $g_{a\gamma}$.
 Furthermore, some observational results 
which can be attributed to the photon-axion conversion might imply its presence~\cite{Aharonian:2007wc, Conlon:2013txa,Moroi:2018vci}.

Here, we propose another way to find the coupling of photon to axion dark matter without using the axion-photon conversion.
The dark matter axion whose field value oscillates around the minimum of its potential 
provides a small difference in the phase velocity between the left-handed and the right-handed photon.
Optical cavity is useful to detect such a small deviation of the phase velocity.
 The birefringence generally caused by ALPs~\cite{Carroll:1989vb} (and specifically by their oscillating background~\cite{Espriu:2011vj}) has been studied,
and recently the authors in \cite{DeRocco:2018jwe} suggested an experiment with a Michelson interferometer.
On the other hand, ring cavity experiments have emerged to test the parity-odd Lorentz violation in the photon sector~\cite{Baynes:2012zz}.
They have measured the variation of the resonant frequency depending on the direction of the light path. 
Similar technique can be applied for our purpose, 
because the resonant frequency of the cavity shifts depends on the polarization of photon, provided that the dark matter axion is coupled to photon.
The dark matter axion predicts the phase velocities of the left and right-handed polarized photon shift with the opposite signs and the same magnitude.
Therefore such shifts of the resonant frequencies of the polarized laser
in the optical cavity are the measurement target in our experiment.
We estimate the reach of our cavity experiment and obtain
the potential sensitivity. 

 This paper is organized as follows.
 In Section 2, we derive the difference in the phase velocity of polarized photons in the presence of axion dark matter and estimate its magnitude.
 In Section 3, we describe the experimental method to probe the axion-photon coupling with our designed cavity.
 In Section 4, we show the potential sensitivity of our experiment and give a short discussion.
 Finally, we conclude the result of this work in Section 5.
 In this paper, we set the unit $\hbar = c = 1$.

\section{II. Phase velocities of photons}

In this section, we present the equations of motions (EoMs) 
for two circular-polarized photons coupled with the axion dark matter and estimate their phase velocities.
We consider the 
axion-photon coupling term
\begin{equation}
\dfrac{g_{a\gamma}}{4}aF_{\mu\nu}\tilde{F}^{\mu\nu} = g_{a\gamma}\dot{a}A_i\epsilon_{ijk}\partial_j A_k + (\text{total derivative}), 
\end{equation}
where dot denotes the time derivative, $a(t)$ is the axion field value, $A_\mu$ is the vector potential, $F_{\mu\nu} \equiv \partial_\mu A_\nu - \partial_\nu A_\mu$, and $\tilde{F}_{\mu\nu} \equiv \epsilon_{\mu\nu\rho\sigma}\partial_\rho A_\sigma/(2\sqrt{-g})$.
Here, we choose the temporal gauge $A_0 = 0$ and the Coulomb gauge $\nabla\cdot \bm{A} = 0$.
 Then the EoM for gauge field reads
\begin{equation}
\ddot{A}_i - \nabla^2A_i + g_{a\gamma}\dot{a}\epsilon_{ijk}\partial_j A_k = 0 \,.
\end{equation}
The present background axion field is written as
\begin{equation}
a(t) = a_0\cos(m t + \delta_\tau(t))\,, 
\label{eq: axion}
\end{equation}
with its constant amplitude $a_0$, its mass $m$ and a phase factor $\delta_\tau(t)$.
In this experiment, we search for the axion dark matter with the mass $m\lesssim 10^{-10}$eV
and the corresponding frequency $f$ is given by
\begin{equation}
f = \dfrac{m}{2\pi} \simeq 2.4\, {\text Hz} \left(\frac{m}{10^{-14}\text{eV}}\right)\,.
\end{equation}
The phase factor $\delta_{\tau}$ can be assumed to be a constant value within the coherence timescale of dark matter $\tau$.

We decompose $A_i$ into two helicity modes with wave number $\bm{k}$ :
\begin{equation}
A_i(t, \bm{x}) = \sum_{\lambda=\pm}\int\dfrac{\dd^3 k}{(2\pi)^3}A^\lambda_{\bm{k}}(t)e^\lambda_i(\hat{\bm{k}})e^{i\bm{k}\cdot\bm{x}} \,,
\end{equation}
where $e^\lambda_i(\hat{\bm{k}}) = e^{\lambda*}_i(-\hat{\bm{k}})$ is the circular polarization vector which obeys $e^{\lambda}_i(\hat{\bm{k}})e^{*\lambda'}_i(\hat{\bm{k}}) = \delta^{\lambda'\lambda},$ and $\epsilon_{ijm}k_je^\pm_m(\hat{\bm{k}}) = \pm k e^\pm_i(\hat{\bm{k}})$.
 Then one finds EoMs for the two polarization modes as
\begin{equation}
\ddot{A}^\pm_k + \omega_\pm^2A^\pm_k = 0\,,
\end{equation}
with
\begin{equation}
\omega_\pm^2 \equiv k^2\left( 1 \pm \dfrac{g_{a\gamma}a_0 m}{k}\sin(m t + \delta_\tau) \right) \label{eq: omega} \,.
\end{equation}
 From \eqref{eq: omega}, we obtain their phase velocities as
\begin{equation}
c_\pm \equiv \dfrac{\omega_\pm}{k} = \left( 1 \pm \dfrac{g_{a\gamma}a_0 m}{k}\sin(m t + \delta_\tau) \right)^{1/2}\,,
\end{equation}
and define their difference as
$\delta c \equiv |c_+ - c_-|$.
The tiny coupling $g_{a\gamma}$ allows us to approximate $\delta c$ by
\begin{equation}
\delta c \simeq \dfrac{g_{a\gamma}a_0 m}{k}\sin(m t + \delta_\tau) \equiv \delta c_0\sin(m t + \delta_\tau)\,.
\end{equation}
 Assuming the laser light with the wavelength $\lambda = 2\pi/k = 1550~\text{nm}$, we can estimate
\begin{equation}
\delta c_0
\simeq 
3\times 10^{-24}\left(\dfrac{g_{a\gamma}}{10^{-12}~\text{GeV}^{-1}}\right)\,,
\end{equation}
where we used the present energy density of the axion dark matter, 
$\rho_a = m^2a_0^2/2 \simeq 0.3~\text{GeV}/\text{cm}^{3}$.


\section{III. Search for axion dark matter using optical ring cavity}

\begin{figure}[tbp]
  \begin{center}
  \includegraphics[width=85mm]{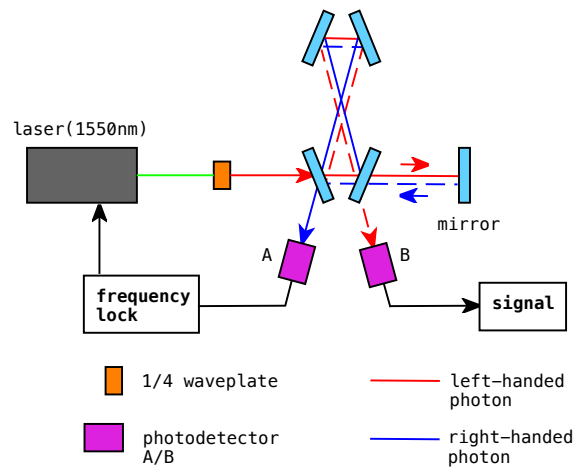}
  \end{center}
  \caption
 {The layout of our double-pass bow-tie cavity.
 The left-handed beam (solid line) is injected to the resonant cavity,
 while the transmit beam reflected by the mirror on the far right goes to the cavity as the right-handed beam (dashed line).
 The photodetector A is used to lock the laser frequency at the resonant  frequency for the injected beam from the left, and the photodetector B monitors
 the modulation of the resonant frequency difference of two optical paths from the beam coming from the right.
 }
 \label{fig: ringcavity}
\end{figure}

In this section, we describe our experiment to detect $\delta c$ caused by the axion dark matter.
The set up of our experiment is schematically illustrated in Figure \ref{fig: ringcavity}.
%
%
First, a laser beam which is circularly polarized by a 1/4 waveplate enters
our bow-tie cavity. For the illustrative purpose, let us assume the incident
beam has the left-handed polarization.
The incident beam to the cavity is partially reflected
by the input mirror and goes to the photodetector A, while the other
part enters the cavity. 
Since the reflection off of a mirror flips the circular-polarization of photon,
the beam changes its polarization each time it is reflected by a mirror.
It should be noted that the beam that enters the cavity from the left has
the right-handed polarization most of the time, because the bow-tie optical
path is stretched in the longitudinal direction.
It eventually goes to either the photodetector A or the mirror on the far right. 
The beam which is reflected from the mirror on the far right 
is partially reflected into the photodetector B
or re-enters the cavity.
Then it has the left-handed polarization most of the time
while traveling inside the cavity in the opposite direction.
Finally some part of the beam goes into the photodetector B.

From 
each photodetector, we can obtain the signal which is  
proportional to the frequency difference between the laser frequency and  the cavity resonant frequency using, for example, Pound-Drever-Hall  
method~\cite{[PDH]}. Using this error signal taken by the photodetector A, the incident laser frequency is stabilized to the resonance of
the (almost) right-handed polarized beam.
We can also obtain the second error signal from the photodetector B, which is proportional to the resonant frequency difference between (almost) left-handed and right-handed beams.
Without the phase velocity modulation $\delta c$ given by the axion dark matter,
the resonant frequency would not depend on the circular-polarizations. 
Therefore our setup works as a null-experiment sensitive to the axion-photon
coupling.

The bow-tie configuration of our optical cavity cancels the Sagnac effect from, for example, the spin of Earth~\cite{Sagnac}.
Most of the environmental noises are also cancelled due to the
double-pass configuration~\cite{Cusack:2002dw}, because the second
error signal observes only the difference in the resonant frequency between the two counter-propagating optical paths in the cavity and their common fluctuations become irrelevant.
The difference of the resonant frequencies between the two optical paths is given by
\begin{equation}
\dfrac{\delta \nu}{\nu} = \dfrac{\delta c}{c} = 
3\times 10^{-24}\left(\dfrac{g_{a\gamma}}{10^{-12}~\text{GeV}^{-1}}\right)
\sin(m t + \delta_\tau) \,, \label{eq: reso}
\end{equation}
and hence the second error signal is expected to be oscillating.
This oscillatory behavior is advantageous for
the signal extraction.

\section{IV. Sensitivity to the axion-photon coupling}

 In this section, we calculate the sensitivity of our experiment to the axion-photon coupling constant.
By virtue of the double-pass configuration, most of the noises from the environmental disturbance are in principle cancelled out by the common mode rejection.
 Then the primary source of noise is the quantum shot noise.
One-sided spectrum of the shot noise of an optical ring cavity is written as \cite{Kimble:2000gu}
\begin{equation}
  \sqrt{S_{\rm{shot}}} = \sqrt{\frac{\lambda}{4 \pi P} \left( \frac{1}{t_r^2} + \omega^2 \right)} \label{eq: shotN} \ ,
\end{equation}
where $\lambda$ is the laser wavelength, $P$ is the input power, and $\omega$ is the angular frequency which is the axion mass $m$ in our case.
 Note that the quantum radiation pressure noise is cancelled out by our double-pass configuration.
 Averaged round-trip time $t_r$ is
\begin{equation}
 t_r = \frac{L \F}{\pi},
\end{equation}
where $L$ is the cavity round-trip length and $\F$ is the finesse.

 If our measurement is limited by the shot noise, the signal-to-noise ratio (SNR) improves with the measurement time $T$ as
\begin{equation}
 (\text{SNR}) = \frac{\sqrt{T}}{2\sqrt{S_{\rm{shot}}}}\frac{\delta c_0}{c}
\end{equation}
as long as the axion oscillation is coherent for $T \lesssim \tau$.
The axion dark matter can be regarded to show a coherent oscillation and $\delta_\tau$ in \eqref{eq: reso} is constant within the coherent time scale $\tau = 2\pi/(mv^2)$. 
 Since the local velocity of dark matter $v$ is around $10^{-3}$, $\tau$ is roughly estimated as
\begin{equation}
\tau \sim 1\,\text{year}\left(\dfrac{10^{-16}\text{eV}}{m}\right) \ .
\end{equation}
 When the measurement time becomes longer than this coherence time $T > \tau$, the phase $\delta_\tau$ is not constant any more and $\delta_\tau$ behaves as a random variable staying constant for each period of $\tau$.
As a consequence, the growth of the SNR with the measurement time changes as
\cite{Budker:2013hfa}
\begin{equation}
(\text{SNR}) 
= \frac{(T\tau)^{1/4}}{2\sqrt{S_{\rm{shot}}}}\frac{\delta c_0}{c} \,.
\end{equation}
%
\begin{figure}[tbp]
  \begin{center}
  \includegraphics[width=85mm]{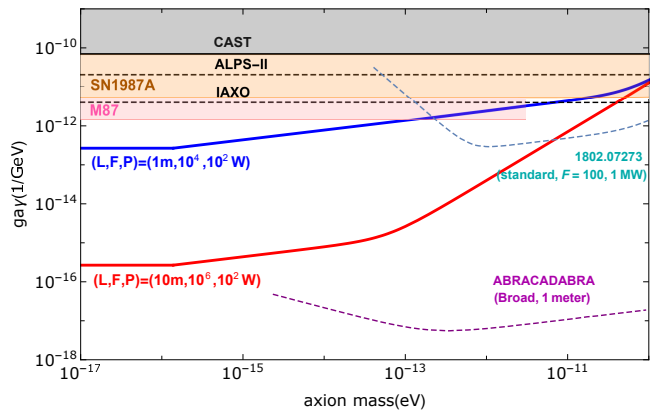}
  \end{center}
  \caption{The sensitivity curves for the axion-photon coupling constant $g_{a\gamma}$ with respect to the axion mass $m$.
 The solid blue (red) line shows the sensitivity of our experiment $(L, F, P) = (1(10)~\text{m}, 10^4(10^6), 10^2(10^2)~\text{W})$.
 The gray band represents the current limit from CAST~\cite{Zioutas:2004hi}.
 The dashed black lines are the prospected limits of IAXO \cite{Vogel:2013bta} and ALPS-II \cite{Ehret:2009sq} missions.
 The dashed turquoise blue and purple lines show the proposed reaches of axion optical interferometer suggested in \cite{DeRocco:2018jwe} and ABRACADABRA magnetometer \cite{Kahn:2016aff}.
 The orange and pink bands denote the astrophysical constraints from the cosmic ray observations of SN1987A \cite{Payez:2014xsa} and radio galaxy M87 \cite{Marsh:2017yvc}.
}
 \label{fig: bound}
\end{figure}
%
Therefore the sensitivity to $\delta c_0 /c$ is limited by
\begin{align}
\frac{\delta c_0}{c}
\lesssim
\begin{cases}
\dfrac{2}{\sqrt{T}} \sqrt{S_{\rm{shot}}} & (T \lesssim \tau) \\
\dfrac{2}{(T\tau)^{1/4}} \sqrt{S_{\rm{shot}}} & (T \gtrsim \tau) \\
\end{cases} \,.
\end{align}
It can be translated into the sensitivity to $g_{a \gamma}$ as
\begin{align}
 g_{a \gamma } &\lesssim 
 \begin{cases}
10^{12} \sqrt{\dfrac{S_{\rm{shot}}}{T}}~[1/\rm{GeV}] & (T \lesssim \tau) \\
10^{12}\sqrt{\dfrac{S_{\rm{shot}}}{(T\tau)^{1/2}}}~[1/\rm{GeV}] & (T\gtrsim \tau) \\
\end{cases}\,.
\end{align}
Figure~\ref{fig: bound} shows the sensitivity of our experiment to the axion-photon coupling constant for different configurations.
Here we set $\lambda=1550$~nm and assumed $T= 1~{\rm year} = 3 \times 10^{7}~{\rm sec}$.
With feasible parameters we can achieve a sensitivity level $g_{a\gamma} \simeq 3 \times 10^{-13} ~\text{GeV}^{-1}$ for $m \lesssim 10^{-16} ~\text{eV}$, which is below the current constraints from axion helioscope experiments, SN1987A and {\sl Chandra} X-ray observations.
Moreover, with more optimistic parameters, our cavity can reach $g_{a\gamma} \simeq 3 \times 10^{-16} ~\text{GeV}^{-1}$ for $m \lesssim 10^{-16} ~\text{eV}$ which will be the best sensitivity among the proposed axion search experiments in this mass range.

 Our optical cavity is made critically-coupled so that most of the beam 
is transmitted and reflected beam power impinging on the photodiode is 
minimized.
 This allows the shot noise limited detection at 100 W input power with current technology.

 We note here that various technical noises at low frequency should be further investigated to determine the sensitivity for lower mass range of the axion.
 Technical noises could be higher than the shot noise especially at low frequencies since technical noises are usually smaller at higher frequencies, and shot noise degrades proportional to frequency above cavity pole (see Eq.\eqref{eq: shotN}). One of the largest technical noises comes from mirror vibration.
 The vibration noise can be estimated with
\begin{equation}
\sqrt{S_{\rm vib}} = \omega^2 A \gamma_{\rm CMRR} \sqrt{S_{\rm seis}} \ ,
\end{equation}
where $A$ is the vibration sensitivity, $\gamma_{\rm CMRR}$ is the common-mode rejection ratio, and $\sqrt{S_{\rm seis}}$ is the ground vibration spectrum.
 Assuming values from cutting-edge technologies, $A=10^{-12}$ /(m/s$^2$), $\gamma_{\rm CMRR}=10^{-4}$ and $\sqrt{S_{\rm seis}}=10^{-9}/f^2 \,{\rm m/\sqrt{Hz}}$, we get $\sqrt{S_{\rm vib}}=4 \times 10^{-24}$.
 This is comparable to the shot noise level below the cavity pole in optimistic cavity parameter case, and reasonable vibration attenuation system would be necessary.

Furthermore, it would be interesting to consider a optical cavity which is sensitive to the axion-photon coupling at higher frequencies, because our cavity does not have much sensitivity at larger mass region $m \gtrsim 10^{-10}$ eV.
 We leave these issues for future work.

\section{V. Conclusion}
\label{Conclusion}
 In this letter, we proposed a novel experiment to probe the coupling of the axion dark matter to photon with a resonant cavity.
We consider the double-pass cavity which aims to detect the difference in the resonant frequencies of the laser beam with the two circular-polarizations. 
Due to the oscillation of the axion dark matter, 
the resonant frequencies are expected to periodically change in time
and we can extract the signal from the irreducible noises.
The sensitivity curve is in principle determined only by quantum shot noise by virtue of the double-pass configuration and hence we can achieve the great sensitivity level for the detection of the axion-photon coupling constant.
In the concrete estimation of the sensitivity, we adopted two sets of parameters, feasible one and optimistic one.
We have demonstrated that both of them can reach sensitivities beyond the current constraints by several orders of magnitude.

\section{Acknowledgement}
\label{Acknowledgement}

 We would like to thank Masaki Ando, Takahiro Tanaka, Hideyuki Tagoshi and Arata Aoki for fruitful discussion and comments.
 In this work, YM and TF are supported by the JSPS Grant-in-Aid for Scientific Research (B) No.~18H01224 and Grant-in-Aid for JSPS Research Fellow No.~17J09103, respectively.


\begin{thebibliography}{99}

\bibitem{Peccei:1977hh} 
  R.~D.~Peccei and H.~R.~Quinn,
  Phys.\ Rev.\ Lett.\  {\bf 38}, 1440 (1977).


\bibitem{Svrcek:2006yi} 
  P.~Svrcek and E.~Witten,
  JHEP {\bf 0606}, 051 (2006)
  [hep-th/0605206];
  A.~Arvanitaki, S.~Dimopoulos, S.~Dubovsky, N.~Kaloper and J.~March-Russell,
  Phys.\ Rev.\ D {\bf 81}, 123530 (2010)
  [arXiv:0905.4720 [hep-th]].


\bibitem{Sikivie:1983ip} 
  P.~Sikivie,
  Phys.\ Rev.\ Lett.\  {\bf 51}, 1415 (1983)
  Erratum: [Phys.\ Rev.\ Lett.\  {\bf 52}, 695 (1984)];
  G.~Raffelt and L.~Stodolsky,
  Phys.\ Rev.\ D {\bf 37}, 1237 (1988).


\bibitem{Hagmann:1998cb} 
  C.~Hagmann {\it et al.} [ADMX Collaboration],
  Phys.\ Rev.\ Lett.\  {\bf 80}, 2043 (1998)
  [astro-ph/9801286];
  S.~Moriyama, M.~Minowa, T.~Namba, Y.~Inoue, Y.~Takasu and A.~Yamamoto,
  Phys.\ Lett.\ B {\bf 434}, 147 (1998)
  [hep-ex/9805026];
  T.~M.~Shokair {\it et al.},
  Int.\ J.\ Mod.\ Phys.\ A {\bf 29}, 1443004 (2014)
  [arXiv:1405.3685 [physics.ins-det]];
  B.~T.~McAllister, G.~Flower, E.~N.~Ivanov, M.~Goryachev, J.~Bourhill and M.~E.~Tobar,
  Phys.\ Dark Univ.\  {\bf 18}, 67 (2017)
  [arXiv:1706.00209 [physics.ins-det]];
  D.~Horns, J.~Jaeckel, A.~Lindner, A.~Lobanov, J.~Redondo and A.~Ringwald,
  JCAP {\bf 1304}, 016 (2013)
  [arXiv:1212.2970 [hep-ph]];
  A.~Caldwell {\it et al.} [MADMAX Working Group],
  Phys.\ Rev.\ Lett.\  {\bf 118}, no. 9, 091801 (2017)
  [arXiv:1611.05865 [physics.ins-det]].


\bibitem{Zioutas:2004hi} 
  K.~Zioutas {\it et al.} [CAST Collaboration],
  Phys.\ Rev.\ Lett.\  {\bf 94}, 121301 (2005)
  [hep-ex/0411033];
  V.~Anastassopoulos {\it et al.} [CAST Collaboration],
  Nature Phys.\  {\bf 13}, 584 (2017)
  [arXiv:1705.02290 [hep-ex]].

\bibitem{Vogel:2013bta} 
  J.~K.~Vogel {\it et al.},
  arXiv:1302.3273 [physics.ins-det].

\bibitem{Ehret:2009sq} 
  K.~Ehret {\it et al.} [ALPS Collaboration],
  Nucl.\ Instrum.\ Meth.\ A {\bf 612}, 83 (2009)
  [arXiv:0905.4159 [physics.ins-det]];
  R.~Bahre {\it et al.},
  JINST {\bf 8}, T09001 (2013)
  [arXiv:1302.5647 [physics.ins-det]];

\bibitem{Betz:2013dza} 
  M.~Betz, F.~Caspers, M.~Gasior, M.~Thumm and S.~W.~Rieger,
  Phys.\ Rev.\ D {\bf 88}, no. 7, 075014 (2013)
  [arXiv:1310.8098 [physics.ins-det]].

\bibitem{Tam:2011kw}
  H.~Tam and Q.~Yang,
  Phys.\ Lett.\ B {\bf 716}, 435 (2012)
  doi:10.1016/j.physletb.2012.08.050
  [arXiv:1107.1712 [hep-ph]].

\bibitem{DeRocco:2018jwe} 
  W.~DeRocco and A.~Hook,
  Phys.\ Rev.\ D {\bf 98}, no. 3, 035021 (2018)
  doi:10.1103/PhysRevD.98.035021
  [arXiv:1802.07273 [hep-ph]].


\bibitem{Sikivie:2013laa} 
  P.~Sikivie, N.~Sullivan and D.~B.~Tanner,
  Phys.\ Rev.\ Lett.\  {\bf 112}, no. 13, 131301 (2014)
  [arXiv:1310.8545 [hep-ph]];

\bibitem{Kahn:2016aff} 
  Y.~Kahn, B.~R.~Safdi and J.~Thaler,
  Phys.\ Rev.\ Lett.\  {\bf 117}, no. 14, 141801 (2016)
  [arXiv:1602.01086 [hep-ph]].


\bibitem{Graham:2015ouw}
  A.~G.~Dias, A.~C.~B.~Machado, C.~C.~Nishi, A.~Ringwald and P.~Vaudrevange,
  JHEP {\bf 1406}, 037 (2014)
  doi:10.1007/JHEP06(2014)037
  [arXiv:1403.5760 [hep-ph]];
  P.~W.~Graham, I.~G.~Irastorza, S.~K.~Lamoreaux, A.~Lindner and K.~A.~van Bibber,
  Ann.\ Rev.\ Nucl.\ Part.\ Sci.\  {\bf 65}, 485 (2015)
  [arXiv:1602.00039 [hep-ex]];
  I.~G.~Irastorza and J.~Redondo,
  arXiv:1801.08127 [hep-ph].



\bibitem{Brockway:1996yr} 
  J.~W.~Brockway, E.~D.~Carlson and G.~G.~Raffelt,
  Phys.\ Lett.\ B {\bf 383}, 439 (1996)
  [astro-ph/9605197];
  J.~A.~Grifols, E.~Masso and R.~Toldra,
  Phys.\ Rev.\ Lett.\  {\bf 77}, 2372 (1996)
  [astro-ph/9606028];

\bibitem{Payez:2014xsa} 
  A.~Payez, C.~Evoli, T.~Fischer, M.~Giannotti, A.~Mirizzi and A.~Ringwald,
  JCAP {\bf 1502}, no. 02, 006 (2015)
  [arXiv:1410.3747 [astro-ph.HE]].


\bibitem{Wouters:2012qd} 
  D.~Wouters and P.~Brun,
  Phys.\ Rev.\ D {\bf 86}, 043005 (2012)
  [arXiv:1205.6428 [astro-ph.HE]];
  D.~Wouters and P.~Brun,
  Astrophys.\ J.\  {\bf 772}, 44 (2013)
  [arXiv:1304.0989 [astro-ph.HE]];
  A.~Abramowski {\it et al.} [H.E.S.S. Collaboration],
  Phys.\ Rev.\ D {\bf 88}, no. 10, 102003 (2013)
  doi:10.1103/PhysRevD.88.102003
  [arXiv:1311.3148 [astro-ph.HE]];
  J.~P.~Conlon, M.~C.~D.~Marsh and A.~J.~Powell,
  Phys.\ Rev.\ D {\bf 93}, no. 12, 123526 (2016)
  [arXiv:1509.06748 [hep-ph]];
  M.~Ajello {\it et al.} [Fermi-LAT Collaboration],
  Phys.\ Rev.\ Lett.\  {\bf 116}, no. 16, 161101 (2016)
  [arXiv:1603.06978 [astro-ph.HE]];
  M.~Berg, J.~P.~Conlon, F.~Day, N.~Jennings, S.~Krippendorf, A.~J.~Powell and M.~Rummel,
  Astrophys.\ J.\  {\bf 847}, no. 2, 101 (2017)
  [arXiv:1605.01043 [astro-ph.HE]];


\bibitem{Marsh:2017yvc} 
  M.~C.~D.~Marsh, H.~R.~Russell, A.~C.~Fabian, B.~P.~McNamara, P.~Nulsen and C.~S.~Reynolds,
  JCAP {\bf 1712}, no. 12, 036 (2017)
  [arXiv:1703.07354 [hep-ph]];

\bibitem{Conlon:2017ofb} 
  J.~P.~Conlon, F.~Day, N.~Jennings, S.~Krippendorf and F.~Muia,
  Mon.\ Not.\ Roy.\ Astron.\ Soc.\  {\bf 473}, no. 4, 4932 (2018)
  [arXiv:1707.00176 [astro-ph.HE]].



\bibitem{Aharonian:2007wc} 
  F.~Aharonian {\it et al.} [H.E.S.S. Collaboration],
  Astron.\ Astrophys.\  {\bf 475}, L9 (2007)
  doi:10.1051/0004-6361:20078462
  [arXiv:0709.4584 [astro-ph]];
  E.~Aliu {\it et al.} [MAGIC Collaboration];
  Science {\bf 320}, no. 5884, 1752 (2008)
  doi:10.1126/science.1157087
  [arXiv:0807.2822 [astro-ph]].
  A.~De Angelis, M.~Roncadelli and O.~Mansutti,
  Phys.\ Rev.\ D {\bf 76}, 121301 (2007)
  doi:10.1103/PhysRevD.76.121301
  [arXiv:0707.4312 [astro-ph]];
  D.~Horns and M.~Meyer,
  JCAP {\bf 1202}, 033 (2012)
  doi:10.1088/1475-7516/2012/02/033
  [arXiv:1201.4711 [astro-ph.CO]].
  M.~Meyer, D.~Horns and M.~Raue,
  Phys.\ Rev.\ D {\bf 87}, no. 3, 035027 (2013)
  doi:10.1103/PhysRevD.87.035027
  [arXiv:1302.1208 [astro-ph.HE]].


\bibitem{Conlon:2013txa} 
  J.~P.~Conlon and M.~C.~D.~Marsh,
  Phys.\ Rev.\ Lett.\  {\bf 111}, no. 15, 151301 (2013)
  doi:10.1103/PhysRevLett.111.151301
  [arXiv:1305.3603 [astro-ph.CO]].
  S.~Angus, J.~P.~Conlon, M.~C.~D.~Marsh, A.~J.~Powell and L.~T.~Witkowski,
  JCAP {\bf 1409}, no. 09, 026 (2014)
  doi:10.1088/1475-7516/2014/09/026
  [arXiv:1312.3947 [astro-ph.HE]].

\bibitem{Moroi:2018vci} 
  T.~Moroi, K.~Nakayama and Y.~Tang,
  Phys.\ Lett.\ B {\bf 783}, 301 (2018)
  doi:10.1016/j.physletb.2018.07.002
  [arXiv:1804.10378 [hep-ph]].

\bibitem{Carroll:1989vb} 
  S.~M.~Carroll, G.~B.~Field and R.~Jackiw,
  Phys.\ Rev.\ D {\bf 41}, 1231 (1990).
  doi:10.1103/PhysRevD.41.1231
  S.~M.~Carroll,
  Phys.\ Rev.\ Lett.\  {\bf 81}, 3067 (1998)
  doi:10.1103/PhysRevLett.81.3067
  [astro-ph/9806099].

\bibitem{Espriu:2011vj} 
  A.~A.~Andrianov, D.~Espriu, F.~Mescia and A.~Renau,
  Phys.\ Lett.\ B {\bf 684}, 101 (2010)
  doi:10.1016/j.physletb.2010.01.005
  [arXiv:0912.3151 [hep-ph]];
  D.~Espriu and A.~Renau,
  Phys.\ Rev.\ D {\bf 85}, 025010 (2012)
  doi:10.1103/PhysRevD.85.025010
  [arXiv:1106.1662 [hep-ph]];


\bibitem{Baynes:2012zz} 
  F.~N.~Baynes, M.~E.~Tobar and A.~N.~Luiten,
  Phys.\ Rev.\ Lett.\  {\bf 108}, 260801 (2012);
  Y.~Michimura, N.~Matsumoto, N.~Ohmae, W.~Kokuyama, Y.~Aso, M.~Ando and K.~Tsubono,
  Phys.\ Rev.\ Lett.\  {\bf 110}, no. 20, 200401 (2013)
  [arXiv:1303.6709 [gr-qc]]; 
  Y.~Michimura, M.~Mewes, N.~Matsumoto, Y.~Aso and M.~Ando,
  Phys.\ Rev.\ D {\bf 88}, no. 11, 111101 (2013)
  [arXiv:1310.1952 [gr-qc]].

\bibitem{[PDH]} 
R. W. P. Drever, J. L. Hall, F. V. Kowalski, J. Hough, G. M. Ford,  
A. J. Munley and H. Ward: Appl. Phys. B 31 (1983) 97.


\bibitem{Sagnac}
G. ac, C. R. Acad. Sci. Paris 157, 708 (1913).

\bibitem{Cusack:2002dw} 
  B.~J.~Cusack, D.~A.~Shaddock, B.~J.~J.~Slagmolen, G.~de Vine, M.~B.~Gray and D.~E.~McClelland,
  Class.\ Quant.\ Grav.\  {\bf 19}, 1819 (2002).


\bibitem{Kimble:2000gu} 
  H.~J.~Kimble, Y.~Levin, A.~B.~Matsko, K.~S.~Thorne and S.~P.~Vyatchanin,
  Phys.\ Rev.\ D {\bf 65}, 022002 (2002)
  [gr-qc/0008026].

\bibitem{Budker:2013hfa} 
  D.~Budker, P.~W.~Graham, M.~Ledbetter, S.~Rajendran and A.~Sushkov,
  Phys.\ Rev.\ X {\bf 4}, no. 2, 021030 (2014)
  doi:10.1103/PhysRevX.4.021030
  [arXiv:1306.6089 [hep-ph]].





\end{thebibliography}
\end{document}